\documentclass[12pt]{iopart}   %% preprint
\usepackage{epsfig}
\usepackage{graphics}

\begin{document}

\title{Density-matrix renormalisation group approach to quantum
impurity problems}

\author{S Nishimoto\dag and E Jeckelmann\ddag}

\address{\dag Fachbereich Physik, Philipps-Universit\"{a}t,
35032 Marburg, Germany}

\address{\ddag Institut f\"{u}r Physik, KOMET 337, Johannes 
Gutenberg-Universit\"{a}t, 55099 Mainz, Germany}

\ead{Eric.Jeckelmann@Uni-Mainz.DE}

\begin{abstract}
A dynamic density-matrix renormalisation group approach to
the spectral properties of quantum impurity problems is presented.
The method is demonstrated on the spectral density of the
flat-band symmetric single-impurity Anderson model.
We show that this approach provides the impurity spectral density
for all frequencies and coupling strengths.
In particular, Hubbard satellites at high energy
can be obtained with a good resolution.
The main difficulties are the necessary discretisation
of the host band hybridised with the impurity and 
the resolution of sharp spectral features such as 
the Abrikosov-Suhl resonance.
\end{abstract}

\pacs{71.27.+a,71.55.Ak,72.15.Qm,78.67.Hc,05.10.Cc}

\vspace{2cm}
%Draft of \today
\submitto{\JPCM}

\maketitle

\section{Introduction}

The single-impurity Anderson model (SIAM)
was invented more than forty years ago to describe  
dilute magnetic impurities in metals~\cite{siam}.
In recent years it has become a generic model for the physics of 
strong local 
electron interactions~\cite{hewson} and is currently attracting much 
interest
as a model for studying quantum dots coupled to leads~\cite{qdots}. 
Its importance has also grown with the advent of the dynamical mean-field
theory (DMFT)~\cite{dmft}. Within the DMFT approach a lattice model
such as the Hubbard model in the limit of high dimensions is mapped onto 
an effective
SIAM which is determined self-consistently.  
Although the properties of the SIAM are generally well understood from 
numerous studies
based on various methods, the situation is not fully satisfactory as far 
as the dynamic 
properties are concerned, especially the zero-temperature high-frequency 
dynamics.
Many-body theories, such as perturbation theory or the local moment 
approach 
(LMA)~\cite{lma1,lma2}, provide an explanative (and often accurate) picture
of quantum impurity dynamics but their accuracy is not known {\it a priori}.
Quantum Monte Carlo (QMC) simulations combined with the Maximum Entropy 
Method~\cite{qmc}
give accurate results at finite temperature but
extrapolations of the dynamic properties to low-temperature are more 
difficult.
The Numerical Renormalisation Group (NRG)~\cite{wilson,nrg,ralph} 
allows one to determine the low-energy dynamics of quantum impurity models 
almost exactly but this method is less precise   
at high-energy because of the necessary logarithmic discretisation of the
host band.

The NRG has been applied to various quantum impurity problems with great 
success but
it has proven more difficult to apply  NRG transformations
to quantum lattice models.
These difficulties provided the motivation for the development of the
density-matrix renormalisation group (DMRG) a decade ago~\cite{steve}.
DMRG is one of the most powerful numerical techniques for studying quantum
lattice many-body systems~\cite{dmrgbook}.
Recently a dynamical DMRG (DDMRG) method has been developed to calculate
dynamic correlation functions at zero temperature in
quantum lattice models~\cite{ddmrg1,ddmrg2}.
Here we extend this approach to the calculation of the spectral 
properties in quantum impurity models. 
The method is demonstrated on the flat-band symmetric SIAM.
We show that DDMRG can provide the spectral density of the SIAM 
for all frequencies and coupling strengths.
In particular, the high-energy spectrum can be determined with a good
resolution. 
Thus the DDMRG approach  is a very useful  
complement to existing techniques for calculating the dynamics of
quantum impurity models.

The Hamiltonian of the SIAM is 
\begin{equation}
\fl \hat{H} =\sum_{k\sigma}
\epsilon_{k} \hat{f}_{k\sigma}^{\dag}\hat{f}_{k\sigma}
+ U \left( \hat{n}_{d\uparrow} -\frac{1}{2}\right)
\left( \hat{n}_{d\downarrow} -\frac{1}{2}\right)
+ \sum_{k\sigma} V_{k}
\left( \hat{f}_{k\sigma}^{\dag}\hat{d}_{\sigma} +
\hat{d}_{\sigma}^{\dag}\hat{f}_{k\sigma}\right) ,
\label{SIAM}
\end{equation}
where $\hat{d}^{\dag}_{\sigma} \ (\hat{d}_{\sigma})$ creates (annihilates)
an electron with spin $\sigma = \uparrow, \downarrow$ in a local level
(the impurity site),
$\hat{n}_{d\sigma} = \hat{d}_{\sigma}^{\dag}\hat{d}_{\sigma}$
and  $\hat{f}^{\dag}_{\sigma} \ (\hat{f}_{\sigma})$ 
creates (annihilates) an electron with spin $\sigma$ in an eigenstate 
of the (non-interacting) host band with dispersion $\epsilon_{k}$.
The sum over the index $k$ runs over all states of the host band.
The hybridisation between the local impurity state and the delocalised
band state $k$ is given by the positive couplings $V_k$.
Electrons in the local level are subject to a Coulomb repulsion $U$.  
(The impurity site potential is set by $\mu = -U/2$ as we will discuss 
the symmetric SIAM only.)

The impurity one-particle Green's function can be written 
($\eta \rightarrow 0^+$)
\begin{equation}
G_{\sigma}(\omega) = \left \langle \hat{d}^{\dag}_{\sigma} 
\frac{1}{\hat{H}-E_0+\omega-i\eta} \hat{d}_{\sigma} \right \rangle
+ \left \langle \hat{d}_{\sigma} 
\frac{1}{E_0-\hat{H}+\omega+i\eta} \hat{d}^{\dag}_{\sigma} \right \rangle ,
\label{green}
\end{equation}
where $E_0$ is the ground state energy
and $\langle \dots \rangle$ represents a ground state expectation value.
In most studies of quantum impurity problems an objective is the 
computation of the impurity spectral density 
\begin{equation}
D_{\sigma}(\omega) = -\frac{1}{\pi} {\rm sgn}(\omega)
{\rm Im} G_{\sigma}(\omega)
= A_{\sigma}(\omega) + B_{\sigma}(\omega)
\label{DOS}
\end{equation}
with 
\begin{eqnarray}
A_{\sigma}(\omega \leq 0) & = & \lim_{\eta \rightarrow 0}
\left \langle \hat{d}^{\dag}_{\sigma}
\frac{\eta}{\pi[(\hat{H}-E_0+\omega)^2+\eta^2]} \hat{d}_{\sigma} \right 
\rangle \label{ADOS}\\
B_{\sigma}(\omega \geq 0) & = & \lim_{\eta \rightarrow 0}
\left \langle \hat{d}_{\sigma} 
\frac{\eta}{\pi[(\hat{H}-E_0-\omega)^2+\eta^2]} 
\hat{d}^{\dag}_{\sigma} \right \rangle  \label{BDOS}
\end{eqnarray}
and $A_{\sigma}(\omega \geq 0) = B_{\sigma}(\omega \leq 0) =0$.
The spectral density fulfills the sum rule 
\begin{equation}
\int_{-\infty}^{\infty} D(\omega) \rmd\omega  =  1 .
\label{sumrule}
\end{equation}

The impurity spectral density of the SIAM model depends on the parameter 
$U$ and  the hybridisation function
\begin{equation}
\Delta(\omega) = \pi \sum_{k} |V_{k}|^2 \delta(\omega - \epsilon_{k}) 
\geq 0 \;  .
\end{equation}
For a symmetric hybridisation function $\Delta(\omega) = \Delta(-\omega)$
the SIAM has a particle-hole symmetry. As consequences,
the Green's function~(\ref{green}) is an odd function of $\omega$ but
the spectral density~(\ref{DOS}) is an even function, 
$D_{\sigma}(\omega) = D_{\sigma}(-\omega)$ [ or
$A_{\sigma}(\omega) = B_{\sigma}(-\omega)]$, and
the Fermi energy remains pinned at $\omega = 0$ for all $U$.
Furthermore, as the total spin is conserved in the SIAM,
$G_{\sigma}(\omega) = G_{-\sigma}(\omega)$.

In the rest of this paper we will explain how 
the spectral density of a quantum impurity problem
such as the symmetric SIAM can be calculated
with DDMRG. To illustrate the various approaches that we have tested
we show results for a well-known particular case of the SIAM:
the flat-band model~\cite{hewson}.
In that case, the host band width is taken to be much larger
than any other energy scales and the hybridisation function
is assumed to be constant on these scales, $\Delta(\omega) = 
\Delta_0$. Our goal is then to compute
$D_{\sigma}(\omega)$ in the relevant energy window
$-W/2 < \omega < W/2$ with $W/2 > U/2, \Delta_0$.
It should be noted that a Friedel sum rule 
\begin{equation}
D_{\sigma}(\omega=0)= \frac{1}{\pi \Delta_0} 
\label{friedel}
\end{equation}
holds at the Fermi energy $\omega = 0$ for all $U \geq 0$ 
in the flat-band symmetric SIAM.
In all numerical results presented here the energy scale 
is set by $\Delta_0 = 1$.

\section{DMRG method for quantum impurity problems}

Since the invention of DMRG there have been numerous applications of this 
method to systems made of one or more impurities coupled to
(generally interacting)  one-dimensional hosts~\cite{olddmrg}.
Recently the growing interest for quantum impurity problems has spurred
the development of DMRG techniques for investigating the more 
relevant problems of impurities coupled
to arbitrary non-interacting host bands both in the context
of quantum dots~\cite{dotdmrg} and of DMFT 
calculations~\cite{dmftdmrg,raas}.

A direct application of the usual DMRG algorithms to the SIAM
Hamiltonian~(\ref{SIAM}) is possible but extremely inefficient
because it includes "long-range" hopping terms (more precisely,
an electron can go from one site to any other site in just two hops).
A better approach is the transformation of the Hamiltonian~(\ref{SIAM})
into a linear chain with nearest-neighbor
interactions only (as in a NRG calculation~\cite{wilson})
\begin{eqnarray}
\hat{H} & = &
V \sum_{\sigma} \left( \hat{c}_{1\sigma}^{\dag}\hat{d}_{\sigma} +
\hat{d}_{\sigma}^{\dag}\hat{c}_{1\sigma}\right)
+ U \left( \hat{n}_{d\uparrow} -\frac{1}{2}\right)
\left( \hat{n}_{d\downarrow} -\frac{1}{2}\right) \nonumber \\
&& +\sum_{j\sigma} a_j \hat{c}_{j\sigma}^{\dag} \hat{c}_{j\sigma}
+ \sum_{j\sigma} \lambda_{j}
\left( \hat{c}_{j\sigma}^{\dag}\hat{c}_{j+1\sigma} +
\hat{c}_{j+1\sigma}^{\dag}\hat{c}_{j\sigma}\right). 
\label{SIAM2}
\end{eqnarray}
The new fermion operators $\hat{c}_{j\sigma}$ corresponds to
electronic states in the host band and are related to the
original representation by a canonical transformation
\begin{equation}
\hat{c}_{j\sigma} = \sum_{k} M_{jk} \hat{f}_{k\sigma} .
\label{transformation}
\end{equation}
The orthogonal matrix $M_{jk}$, the diagonal terms
$a_j$ and the nearest-neighbor
hopping terms $\lambda_j$ are calculated with
the Lanczos iterative algorithm for tridiagonalising
a symmetric matrix starting from the initial vector
\{$M_{1,k}  =  V_k / V$\} with $V^2= \sum_{k} V_k^2$. 
This calculation must be carried out with very high
numerical accuracy (quadruple or higher precision for
floating-point operations) but does not present any other difficulty.
If the original Hamiltonian~(\ref{SIAM}) is particle-hole
symmetric, the diagonal terms $a_j$ vanish.

\begin{figure}
\begin{center}
\resizebox{10cm}{!}{\epsfbox{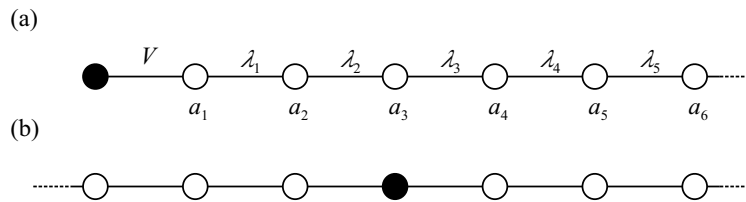}}
\end{center}
\caption{\label{fig1} One-dimensional lattice configurations for
applying a DMRG algorithm to quantum impurity problems:
(a) Impurity site (solid circle) at one end of the chain representing
the host band (open circles). 
(b) Impurity site in the middle of two chains.}
\end{figure}

The Hamiltonian~(\ref{SIAM2}) describes an impurity coupled
to one end of a one-dimensional chain representing the host band states
(see figure~\ref{fig1}).
Here we will use only this configuration. 
However, one can easily imagine different configurations such as an impurity
site located between two chains as also shown in figure~\ref{fig1}.
In that case, the left and right chains could correspond to band states
below ($\epsilon_k < 0$) and above ($\epsilon_k > 0$) the Fermi energy,
respectively. Another possibility is that the left and right chains
could correspond to up-spin and down-spin band states, respectively.
The transformation~(\ref{transformation})
described above can easily be adapted for either
cases. It can also be generalised to more complicated impurity problems
with more than one local level or more than one host band.

The finite-system DMRG algorithm~\cite{steve, dmrgbook}
can be used to calculate ground state properties of the
Hamiltonian~(\ref{SIAM2}). In particular, the ground state wavefunction
$|\Psi_0\rangle$ and the ground state energy $E_0$ can readily be obtained.
To compute dynamic properties such as the 
impurity Green's function~(\ref{green})
we use the dynamical DMRG~\cite{ddmrg2}.
This approach is based on a variational principle.
One can easily show that for $\eta > 0$ and a fixed frequency
$\omega$ the minimum of the functional
\begin{equation}
\fl W(\Psi) =
\left \langle \Psi \left | \left(E_0+\omega-\hat{H}\right)^2 +\eta^2
\right | \Psi 
\right \rangle
 + \eta \left \langle \Psi_0 \left | \hat{d}_{\sigma} \right | 
 \Psi \right \rangle
+ \eta  \left \langle \Psi \left | \hat{d}_{\sigma}^{\dag} 
\right | \Psi_0  \right \rangle
\label{functional}
\end{equation}
with respect to all quantum states $|\Psi\rangle$ is
\begin{equation}
W(\Psi_{\rm min}) = 
\left \langle \Psi_0 \left | \hat{d}_{\sigma} 
\frac{-\eta^2}{\left(E_0+\omega-\hat{H}\right)^2 +\eta^2}
\hat{d}_{\sigma}^{\dag} \right | \Psi_0 \right \rangle .
\end{equation} 
The functional minimum is related  to the convolution 
of the spectral density~(\ref{BDOS}) with 
a Lorentz distribution of width $\eta$ by
\begin{equation}
W(\Psi_{\rm min}) = -\pi \eta B_{\sigma}^{\eta}(\omega).  
\end{equation}
A similar result is obtained for the spectral density~(\ref{ADOS}) 
if one substitutes $\hat{d}_{\sigma}$
for $\hat{d}^{\dag}_{\sigma}$, $-\omega$ for $\omega$ and 
$A_{\sigma}^{\eta}(\omega)$ for $B_{\sigma}^{\eta}(\omega)$
in the above equations.

The DDMRG method consists essentially in  
minimising the functional~(\ref{functional}) numerically using
the finite-system DMRG algorithm.
Thus DDMRG provides the spectral densities
$A^{\eta}_{\sigma}(\omega)$ and $B^{\eta}_{\sigma}(\omega)$
for a finite broadening $\eta$.
The full spectral density~(\ref{DOS}) convolved with the Lorentz
distribution 
\begin{equation}
D^{\eta}_{\sigma}(\omega) =  \int_{-\infty}^{\infty} \rmd\omega'
D_{\sigma}(\omega') \frac{\eta}{\pi[(\omega-\omega')^2+\eta^2]}
\label{convolution}
\end{equation}
is given by the sum of  $A^{\eta}_{\sigma}(\omega)$ and 
$B^{\eta}_{\sigma}(\omega)$.
The real part of the Green's function can be calculated 
with no additional computational cost but is generally less accurate.
The necessary broadening
of spectral functions in DDMRG calculations
is actually very useful for studying continuous spectra
or doing a finite-size scaling analysis~\cite{ddmrg2}.

The objective of our DMRG simulations is to obtain
physical quantities with a sufficient accuracy at the lowest possible
computational cost. For this purpose, the lattice configuration used
in the present work makes necessary an adaptation of the usual DMRG 
principles for measurements.
In standard DMRG calculations the one-dimensional lattice is split
in two blocks of sites separated by two intermediate 
sites~\cite{steve,dmrgbook}. 
Calculations are carried out in an effective Hilbert space of dimension
$D \approx n^2 m_L m_R$,
where $n$ is the number of states per site ($n=4$ counting the spin 
degeneracy in our case) and $m_{L,R}$ are the numbers of states
used to describe the left and right blocks, respectively. 
Note that $m_{L,R} \approx \min(m,4^{N_{L,R}})$, where
$m$ is the maximum number of density-matrix eigenstates kept 
and $N_L$ and $N_R$ are the numbers of sites in the left and right blocks, 
respectively.
The DMRG method errors diminish (often exponentially fast) as
$m$ is increased while the computational cost increases as
$m^3$ (at least in theory).
During a DMRG simulation the position where the chain is split is moved
repeatedly through all sites and the measurement precision 
for a fixed number $m$ varies with that position.
Measurements of local quantities such as the density 
$\hat{c}^{\dag}_{j\sigma} \hat{c}_{j\sigma}$ are most accurate
if the site $j$ is one of the two intermediate
sites. Measurements of global quantities, such as
the ground state energy or long-range correlations, are most accurate
if the chain is split in two blocks of equal size because the
effective Hilbert space  dimension $D$ is maximal then.
In the present application we have found that it is necessary to compromise
in order to optimize the ratio between accuracy and computational cost
in calculations of the spectral density for an impurity
located at a chain end.
Measurements have to be done as close as possible to the impurity but
far enough from the chain end for the effective Hilbert space
to approach its maximal dimension $D = n^2 m^2$.
This implies that measurements are to be done when the left
block has reached the size of $N_L \approx \ln(m)/\ln(4)$ sites
(assuming the impurity to be at the left chain end).

In practice, we keep enough density-matrix eigenstates to make (D)DMRG
truncation errors negligible (we have used up to
$m= 800$ states in the present work). Thus our DDMRG  
results for finite systems are numerically exact in the same sense as 
``exact diagonalisation" results are.
The main source of errors are finite-size effects, which corresponds to 
the discretisation of the continuous host band in the SIAM and are 
discussed in the next section.

\begin{figure}
\begin{center}
\resizebox{10cm}{!}{\epsfbox{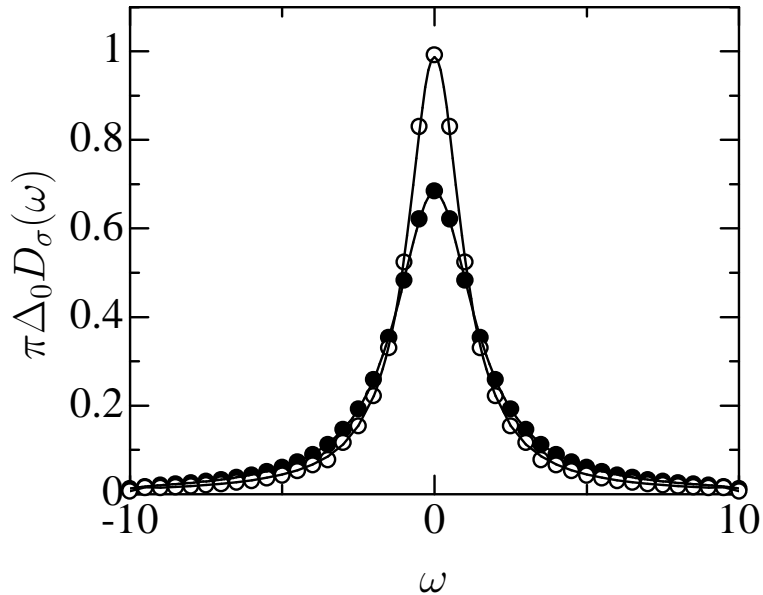}}
\end{center}
\caption{\label{fig2} Spectral density at $U=0$
calculated with a constant host band discretisation (solid circles)
for $W=20\Delta_0, N=59, \Delta\epsilon\approx 0.34\Delta_0$, and
$\eta=0.5$,
then deconvolved (open circles). Solid lines show the exact results
(\ref{lorentzian})
without broadening and with $\eta=0.5$ for comparison. 
}
\end{figure}

In figure~\ref{fig2} we show the spectral density of the
flat-band symmetric SIAM calculated with DDMRG for $U=0$. 
The exact spectral density is a Lorentzian  of width $\Delta_0$.
With the broadening~(\ref{convolution}) it becomes
\begin{equation}
D^{\eta}_{\sigma}(\omega) = \frac{\Delta_0 + \eta}
{\pi[\omega^2+(\Delta_0+\eta)^2]} .
\label{lorentzian}
\end{equation}
On the scale of figure~\ref{fig2} there is no visible difference 
between our numerical results and this exact result, which demonstrates
the accuracy of our method.
Note that the $U=0$ limit is a relevant accuracy test. 
The hybridisation $\Delta(\omega)$ with the host band is 
renormalised in the density-matrix renormalisation and thus
our method could become exceptionally accurate 
for some specific hybridisations [for instance, $\Delta(\omega)=0$]
but the local Coulomb interaction $U$ is always treated exactly
and thus does not affect the method accuracy directly.

With DDMRG the computational cost (memory and processor time)
of computing $D_{\sigma}^{\eta}(\omega)$ for a single frequency $\omega$
is proportional to the number $N$ of sites in the system.
In practice, the total processor time for calculating the full spectrum
grows as $\sim N^3$ because one usually reduces $\eta$ and increases the 
frequency resolution as $N$ is increased.

\section{Discretisation of the host band}

We are interested in the properties of the 
SIAM with a continuous host band and a continuous 
hybridisation function $\Delta(\omega)$ but DMRG calculations
can be performed on finite lattices only.
Therefore, we must discretise the host band and carry out DMRG
calculations for a finite number $N$ of host band eigenstates
corresponding to energies $\epsilon_k (k=1, \dots, N)$, 
then extrapolate the results to a continuous host band
($N \rightarrow \infty$).
Choosing an appropriate discretisation of the host band
(i.e., selecting the $N$ band state energies $\epsilon_k$)
turns out to be the greatest difficulty in applying
DDMRG to the SIAM.

The spectral density~(\ref{ADOS}) [or similarly~(\ref{BDOS})]
can be written in the Lehmann representation \hfill
\begin{equation}
A_{\sigma}(\omega) = \sum_{n} \left | \left \langle \Psi_n \left |
\hat{d}_{\sigma}\right |\Psi_0 \right \rangle \right |^2
\delta(\omega + E_n -E_0) ,
\label{lehmann}
\end{equation}
where $| \Psi_n \rangle$, $n=0,1, \dots$, design the eigenstates
of $\hat{H}$ contributing to $A_{\sigma}(\omega)$
(i.e., with non-vanishing matrix elements) and $E_n$
their respective energies.
On a finite lattice this spectrum is always discrete.
To determine the exact SIAM spectral density for $N\rightarrow \infty$
it is necessary to use a broadening $\eta$ larger than the space
$E_{n+1}-E_n$
between two consecutive eigenstates in 
dense parts of the spectrum 
(for instance, in continuous parts of the spectrum)~\cite{ddmrg2}. 
In the flat-band SIAM 
(and also in self-consistent SIAM derived in DMFT calculations for 
the Hubbard model~\cite{dmftdmrg}) the distribution of excited 
states contributing to the spectral density for finite $N$
is essentially determined by the distribution of selected energies
$\epsilon_k$.
Therefore, we must
use 
\begin{equation}
\eta  > E_{n+1}-E_n \sim \epsilon_{k+1} - \epsilon_k.
\label{broadening}
\end{equation}
In other words, the host band discretisation directly limits the resolution
of spectral density calculations.
Note that this is a general difficulty for all methods based
on a discretisation of the SIAM, not only for DDMRG calculations.
For instance, a similar problem limits the resolution of NRG
calculations at high energy.

In practice, we choose $N$ values $\epsilon_k$ in the relevant
energy window $|\omega| < W/2$ and calculate the
corresponding hybridisation terms 
\begin{equation}
V_k^2 = \frac{1}{\pi}
\int_{\delta_k}^{\delta_{k+1}} \Delta(\omega) \rmd\omega ,
\end{equation}
where $\delta_k = (\epsilon_{k-1} + \epsilon_k)/2$.
If $\Delta(\omega)$ varies slowly for $\omega \approx \epsilon_k$ 
on the scale $\Delta\epsilon_k = (\epsilon_{k+1} - \epsilon_{k-1})/2$ 
then $ V_k^2 \approx \Delta(\omega=\epsilon_k) \Delta\epsilon_k/\pi$.
For the flat-band SIAM that we are considering here this last equation
is obviously exact.
To preserve the particle-hole symmetry in the discrete SIAM
we only include pairs of band states with opposite energies
in the $N$ selected values.
For even $N$ this means that there are $N/2$ different pairs
($\epsilon_k,-\epsilon_k$), while for odd $N$ there are $(N-1)/2$ 
such pairs and a state at $\epsilon_k=0$.
Including the impurity site, the total number of sites
in the lattice is thus $N+1$.
The ground state contains an equal number of electrons and has
a minimal total spin $S=0$ for odd $N$ and $S=1/2$ for even $N$.
Note that in that last case ($S \neq 0$) we need to compute
the spectral density for both $\sigma=\uparrow$ and $\sigma=\downarrow$
and take the average, as the spin-flip symmetry $\sigma \leftrightarrow
-\sigma$ is broken. It is thus more efficient to work
with an odd number $N$ of host band states.

\begin{figure}
\begin{center}
\resizebox{10cm}{!}{\epsfbox{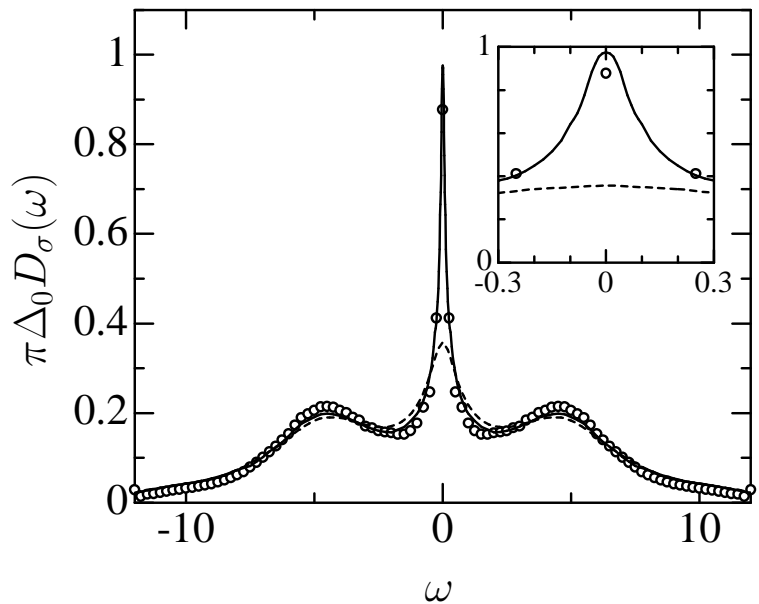}}
\end{center}
\caption{\label{fig3} Spectral density for $U=2.5\pi\Delta_0$
and $W=24\Delta_0$.
Calculated for a constant discretisation 
$\Delta\epsilon\approx 0.34\Delta_0$ and  $\eta = 0.5\Delta_0$ (dashed
line), calculated using variable discretisation  
$0.2 \geq \Delta\epsilon/\Delta_0 \geq 0.0059$ and 
broadening $0.25 \geq \eta/\Delta_0
\geq 0.01$ (solid line), and calculated using a variable discretisation
$0.76 \geq \Delta\epsilon/\Delta_0 \geq 0.16$
and a constant broadening $\eta=0.25 \Delta_0$ then deconvolved 
(circles). Inset: expanded view around the Fermi level $\omega=0$.}
\end{figure}

The simplest discretisation scheme consists in choosing $N$ equidistant
energies $\epsilon_k$  ($\Delta\epsilon_k = \Delta\epsilon 
\approx W/N$) in the relevant energy window
($W/2 > |\epsilon_k|$).
In that case we use a  constant broadening $\eta \approx \Delta\epsilon$.
This approach has been used for the $U=0$ results shown in 
figure~\ref{fig2}. 
A constant discretisation is sufficient provided that the spectrum does
not contain any structure with a width smaller than
$\Delta\epsilon \propto 1/N$, which is readily achieved in the 
weak-coupling SIAM.
It is well-known~\cite{hewson,lma1,ralph} 
that for intermediate to strong couplings $U$ 
the broad spectral feature around $\omega =0$ shrinks to a sharp
peak (the so-called Abrikosov-Suhl resonance). The spectral
weight is progressively transfered to two Hubbard satellites
around $\omega \approx \pm U/2$ for increasing $U$.
For instance, in figure~\ref{fig3} we show the spectral density for
the intermediate coupling $U=2.5\pi\Delta_0$ 
calculated with a constant discretisation 
$\Delta\epsilon\approx 0.34\Delta_0$ and a broadening $\eta = 0.5\Delta_0$.
This spectrum agrees qualitatively with LMA and NRG
results~\cite{lma1,ralph}.
Quantitatively, however, the DDMRG spectrum is obviously inaccurate,
especially around $\omega \approx 0$.
For instance, at the Fermi level
the Friedel sum rule~(\ref{friedel}) 
is clearly not fulfilled.
As for the $U=0$ case this apparent discrepancy is due to the broadening
$\eta$.

\begin{figure}
\begin{center}
\resizebox{10cm}{!}{\epsfbox{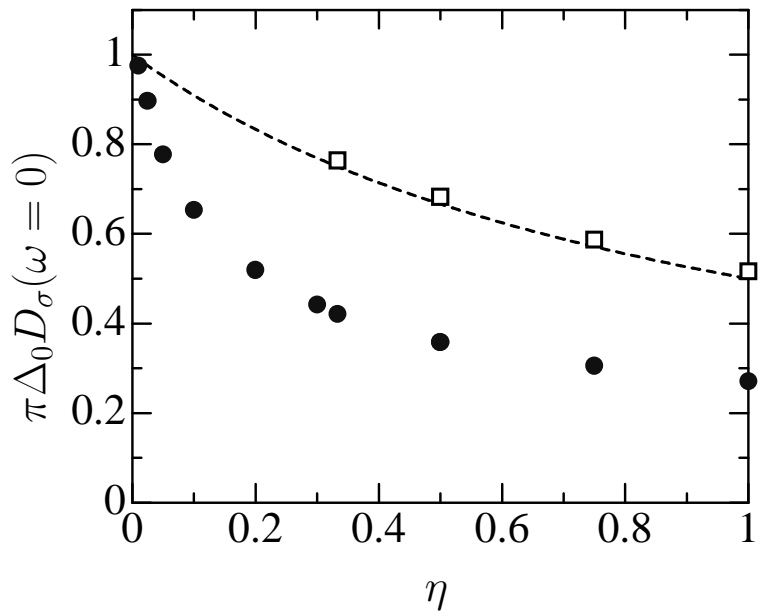}}
\end{center}
\caption{\label{fig4} Fermi level spectral density
$\pi\Delta_0 D_{\sigma}^{\eta}(\omega=0)$
as a function of the broadening $\eta$.
Exact result for $U=0$ (dashed line) and DDMRG results for
$U=0$ (squares) and $U=2.5\pi\Delta_0$ (circles).}
\end{figure}

To obtain better results one can perform calculations with
a higher resolution (which means smaller $\eta \approx \Delta\epsilon
\sim 1/N$ and thus larger lattice size $N$) and possibly extrapolate to
the limits $\eta \rightarrow 0$ and $N \rightarrow \infty$.
In figure~\ref{fig4}
we show $D^{\eta}_{\sigma}(\omega = 0)$ calculated with DDMRG at
$U=2.5\pi \Delta_0$ for several values of $\eta$.
For comparison we also show the exact scaling for the non-interacting
($U=0$) case. 
Clearly, $D^{\eta}_{\sigma}(\omega = 0)$ tends to the exact result as
$\eta \rightarrow 0$ but the convergence is slow and will
become worse for sharper resonances (larger $U$).
For instance, for $U=8\pi\Delta_0$ we can already clearly observe 
a very sharp Abrikosov-Suhl resonance using a resolution of
$\Delta\epsilon \approx \eta = 10^{-4} \Delta_0$ but
the height of the peak is only $D^{\eta}_{\sigma}(\omega=0)
\approx 0.06/ \pi\Delta_0$.
Actually, one can guess that a broadening $\eta$ smaller than
the Kondo scale $\propto \exp(-\pi U/8\Delta_0)$~\cite{hewson,lma1} 
is required to obtain the low-energy spectral density
with a good resolution and accuracy.
Therefore, the required system size $N \sim \eta^{-1}
\propto \exp(\pi U/8\Delta_0)$ increases 
exponentially with $U$ if one uses a constant discretisation.
This approach is clearly not practicable.

\begin{figure}
\begin{center}
\resizebox{10cm}{!}{\epsfbox{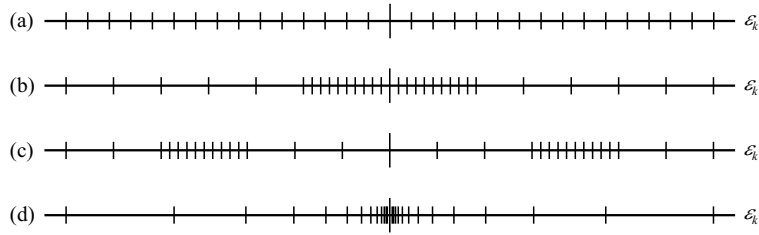}}
\end{center}
\caption{\label{fig5} Various host band discretisation schemes:
(a) constant, (b) variable with a higher resolution
at low energy, (c) variable with a higher resolution
at intermediate energy, and (d) logarithmic.}
\end{figure}

As an interest of many quantum impurity problems 
is the investigation of sharp resonances in the spectral density 
and a analogous sharp quasi-particle peak
is an essential feature of DMFT calculations,
it is desirable to achieve a high accuracy  and resolution
for these spectral features too.
A better approach is the use of a variable discretisation.
For instance, one can select two sets of equidistant energies $\epsilon_k$
in the host band with a energy resolution 
$\Delta\epsilon$ for high energies 
$W/2 > | \epsilon_k | > W^*/2$ and
with another energy resolution $\Delta\epsilon' < \Delta\epsilon$ for 
$| \epsilon_k | < W^*/2$ (see figure~\ref{fig5}).
In that case the broadening $\eta$ also depends on $\omega$.
Typically, we use $\eta \approx \Delta\epsilon$
for $W/2 > | \omega | > W^*/2$ and
$\eta \approx \Delta\epsilon'$ for $| \omega| < W^*/2$.
This allows us to obtain the spectral density around the 
Abrikosov-Suhl resonance 
with a much higher resolution for a given number of 
host band sites $N$. 
It is also possible to use more than two different 
energy resolutions, to use a higher energy resolution 
for the high energy sector than for the low energy sector, or even to 
select a higher resolution at intermediate energies (see figure~\ref{fig5}).
Moreover, it is possible to combine the results of the high-resolution
sectors obtained with different variable discretisations
in a single high resolution spectrum.

This approach allows us to achieve a resolution
which is significantly better than
with the constant discretisation for the same number of
host band states $N$.
For instance, figure~\ref{fig3} shows the spectral density
obtained with variable discretisation $\Delta\epsilon$
and broadening $\eta$
ranging from  $\Delta\epsilon = 0.2 \Delta_0$
($\eta = 0.25 \Delta_0$) for the Hubbard satellites
to  $\Delta\epsilon = 0.0059 \Delta_0$ ( $\eta =0.01 \Delta_0$)
for $\omega \rightarrow 0$.
This approach clearly gives much better results
for the sharp Abrikosov-Suhl resonance than the constant
discretisation approach.
The Friedel sum rule~(\ref{friedel}) at the Fermi level
is fulfilled within 3 \%.
It should be noted that this approach is not without problem.
The most obvious drawback is that a variable broadening
$\eta=\eta(\omega)$ breaks the sum rule~(\ref{sumrule})
and can significantly change the spectrum shape if 
it varies faster than the bare spectrum as a function of $\omega$.
(This could be the case for the Abrikosov-Suhl resonance
shown as a solid line in the inset of figure 3.)

To take advantage of the higher resolution provided by
a variable discretisation while
keeping the benefit of a constant broadening $\eta$
we have devise the following scheme.
The relevant energy window is split into 
several intervals. For each interval, a DDMRG calculation
is done using a variable discretisation with a high resolution
in that interval and a lower resolution outside.
The spectral function is calculated for frequencies in that
interval.
Then the results obtained in each calculation are combined
into a full spectrum.
The computational effort is equal to that of a single
calculation with  a constant discretisation for
the same number of host band states, but the resolution is higher.
In figure~\ref{fig3} we show the spectral density calculated 
using this scheme with a high resolution of $\Delta_\epsilon
= 0.16 - 0.20 \Delta_0$ inside each interval and a lower 
resolution of $\Delta_\epsilon = 0.71 - 0.76 \Delta_0$ outside. 
The spectrum has 
been calculated with $\eta=0.25 \Delta_0$ then deconvolved 
(see next section).
This spectrum is clearly sharper than the one obtained with a
constant discretisation and, by comparison with
related NRG and LMA results, it seems also more accurate.
The improvement is due both to the lower broadening $\eta$
and the deconvolution.
Nevertheless, the variable discretisation scheme with varying broadening
$\eta$ is more accurate for low frequency  and 
gives a much sharper Abrikosov-Suhl resonance.

As in NRG calculations~\cite{wilson}
a logarithmic discretisation of the host band 
$\epsilon_k = (W/2)\Lambda^{-k}$ (with $\Lambda > 1$ and $
k=1,2,\dots,N/2$) is possible.
Ground state DMRG calculations can easily been performed
although the computational effort is significantly greater than with NRG.
Calculating the spectral density with DDMRG is more problematic.
The main problem is that we have not found
any satisfactory method to broaden the spectrum.
In NRG calculations~\cite{ralph} the spectrum is convolved
with a function which vanishes for $\omega \rightarrow 0$ and
broadens the spectrum on a logarithmic scale $\eta \propto \omega
\sim \Delta\epsilon_k$.
In DDMRG calculations we always get the spectrum convolved 
with a Lorentzian~(\ref{convolution}).
To imitate the logarithmic broadening we have tried a
"Lorentzian" broadening~(\ref{convolution}) with
a variable $\eta =\eta(\omega') = \lambda \omega'$.
Note that with the DDMRG method 
we can change $\eta$ as a function of $\omega$
but not of $\omega'$ in~(\ref{convolution}).
Fortunately, if $G_{\sigma}(\omega)$ is
the impurity Green's function obtained with DDMRG for
$\eta(\omega) =\lambda\omega$, the desired
Green's function $\tilde{G}_{\sigma}(\omega)$ 
with $\eta(\omega') = \lambda \omega'$ can be obtained through
\begin{equation}
\tilde{G}_{\sigma}(\omega) = \frac{1+i\lambda}{1+\lambda^2} \
G_{\sigma} \left (\frac{\omega}{1+\lambda^2} \right ) .
\end{equation}
Note that the condition~(\ref{broadening}) implies
$\lambda > 1-\Lambda^{-1}$.
We have tested this approach on the flat-band SIAM for
various values of $U$.
As expected, the high-energy spectrum is widely broadened and
qualitatively similar to corresponding NRG results~\cite{ralph}.
In the low-energy spectrum, however, this scheme clearly does not
work because $D_{\sigma}(\omega)$ seems to diverge 
for $\omega \rightarrow 0$ or at least the Fermi level spectral density
$D_{\sigma}(0)$ is much larger than the exact value $1/(\pi\Delta_0)$. 
We think that this failure is due to the broadening with the
Lorentzian function~(\ref{convolution}), which does not vanish
for $\omega\rightarrow 0$ contrary to the function used
in NRG calculations~\cite{ralph}.
Therefore, a logarithmic discretisation does not seem to be useful 
for DDMRG calculations.
It should be noted that with the Lanczos-DMRG~\cite{karen} or the
correction-vector DMRG~\cite{till} it should be possible to calculate
the spectral density in the Lehmann representation~(\ref{lehmann})
(i.e., without broadening) and to broaden it on a logarithmic scale 
as in NRG calculations.
This approach could give better results for the low-energy spectrum
than the DDMRG approach discussed here but has yet to be tested.

In summary, the host band discretisation determines the resolution of
DDMRG calculations for the impurity spectral density.
No single discretisation scheme is appropriate for all cases.
Nevertheless, a combination of different schemes can
be used for the various features of the same spectrum.
Therefore, DDMRG allows one to calculate
a spectral density with high resolution for all frequencies provided 
the host band discretisation 
and the broadening are adapted to the specific problem and its spectral 
features.

\section{Deconvolution of DDMRG spectra}

An approach for obtaining sharper spectra is the deconvolution of 
the DDMRG data.
In theory, a deconvolution means solving (\ref{convolution}) for 
$D(\omega)$ using the DDMRG data in the left-hand side.
However, this is typically a
ill-conditioned inversion problem~\cite{numrep}
which generally cannot be solved numerically.
Moreover, if it was possible to do this calculation exactly,
one would obtain the discrete spectral density of the SIAM
on a finite lattice of $N+1$ sites.

Nevertheless, the broadened spectral density of the SIAM 
on a infinite lattice ($N \rightarrow \infty$)
is usually almost identical to the spectral density of the discretised
SIAM ($N < \infty$).
For instance, for $U=0$ there is no visible difference
in figure~\ref{fig2} between the DDMRG results for $N=59$
and the exact result for $N\rightarrow \infty$.
[A necessary condition for a large enough $\eta$ is
(\ref{broadening}).]
Therefore, one can make the approximation that DDMRG data for 
$D_{\sigma}^{\eta}(\omega)$ describe the broadened spectral density
for $N \rightarrow \infty$ and solve
(\ref{convolution}) approximately under the condition
that $D_{\sigma}(\omega)$ is the exact spectral density of the 
SIAM.
For instance, one can require that $D_{\sigma}(\omega)$ is a continuous
and relatively smooth function.

In practice, we calculate $D_{\sigma}(\omega)$ using the algorithm
presented in~\cite{dmftdmrg} or using standard linear 
regularisation methods for the inverse problem~\cite{numrep}.
Other possibilities include using Fourier transformations~\cite{raas}
or the
Maximum Entropy Method as in QMC simulations~\cite{qmc}.
In figure~\ref{fig2} one sees that the deconvolved DDMRG spectral density
agrees perfectly with the exact result for $U=0$.
Similarly, a deconvolution gives excellent results for finite
but weak coupling $U$.
For $U=2.5\pi\Delta_0$ 
the deconvolved spectral density in figure~\ref{fig3} shows sharper
Abrikosov-Suhl resonance and Hubbard satellites 
than the original DDMRG data for $\eta=0.25\Delta_0$ (not shown).
However, the Friedel sum rule~(\ref{friedel}) at the Fermi level
is far from being fulfilled (the relative error is about 10 \%), which
indicates that the Fermi level resonance is still too broad despite
the deconvolution.
In general, we have found that the deconvolution  works very well
for the Hubbard satellites and other broad structures
(broader than $\eta$) but bring only a partial improvement for sharp
peaks.

\begin{figure}
\begin{center}
\resizebox{10cm}{!}{\epsfbox{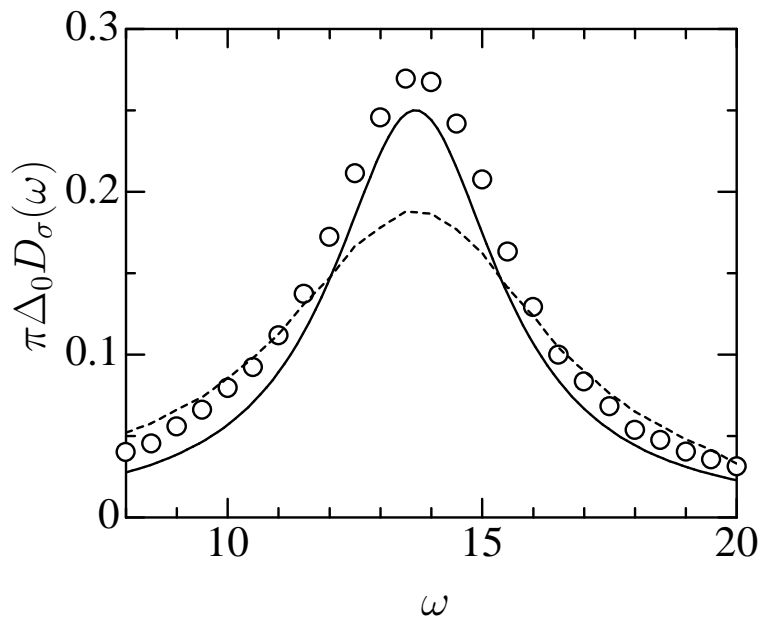}}
\end{center}
\caption{\label{fig6} Upper Hubbard satellite for
$U=8\pi\Delta_0$. Calculated with DDMRG for $W= 40 \Delta_0$
and a constant discretisation $\Delta\epsilon = 1.29 \Delta_0$
and $\eta = \Delta_0$ (dashed line) then deconvolved 
(circles). The solid line shows the LMA result~\cite{lma1}
for $U \gg \Delta_0$ (see text). 
}
\end{figure}

Deconvolution methods become rapidly instable if a variable
broadening $\eta$ is used in (\ref{convolution}).
Therefore, we apply them to spectra calculated  with a constant 
$\eta$ only.
Nevertheless, if the spectrum is made of well separated structures
(as the Abrikosov-Suhl peak and the Hubbard satellites are
for strong coupling $U \gg \Delta_0$),
it is possible to deconvolve each structure separately.
In that case we can use a different $\eta$ for each structure. 
In figure~\ref{fig6} we show the upper Hubbard satellite
calculated with DDMRG for $U=8\pi\Delta_0$ then deconvolved.
Our numerical result agrees remarkably well with the LMA prediction for
this spectral structure in the strong coupling regime $U \gg \Delta_0$: 
a Lorentz distribution of width $2\Delta_0$ and total weight $\frac{1}{2}$
centered at $\omega_c = U/2$. 
Note that in figure~\ref{fig6} we have chosen $\omega_c = 13.7 \Delta_0$
to fit our data rather than the value $\omega_c = U/2 = 4\pi\Delta_0
\approx 12.6 \Delta_0$. 
This shift of $1.1 \Delta_0$ is due to 
corrections to $\omega_c$ of the order of $\Delta_0/U$ \cite{lma1}
and the finite discretisation $\Delta\epsilon= 1.29 \Delta_0$ used 
in our calculation.
The deconvolved DDMRG spectral density appears to contain more
spectral weight than the LMA result because we have used a rather
narrow energy window $|\omega| < W/2 = 20 \Delta_0$ in the DDMRG
calculation and thus the spectral weight which normally lies 
in the high energy tail ($|\omega| > W/2$) of the Lorentz distribution 
has been shifted to low energy ($|\omega| < W/2$).  

\section{Conclusion}

We have extended the DDMRG method~\cite{ddmrg2}
to the investigation of dynamic
properties in quantum impurity problems.
The method has been demonstrated on the flat-band  symmetric SIAM.
We have obtained accurate results for the impurity spectral density
for all frequencies and coupling strengths. The main difficulty
of this approach is the choice of an appropriate discretisation
of the host band to resolve sharp spectral features such as the
Abrikosov-Suhl resonance in the strong-coupling regime of the SIAM.

Our method can readily be used to study quantum lattice 
many-body systems in the framework of the DMFT.
We have already applied this DDMRG-DMFT technique to the
Hubbard model in infinite dimensions. For the metallic phase
at weak coupling $U$ our numerical results have a better
resolution than those obtained with NRG and  are in excellent agreement
with weak-coupling perturbation theory~\cite{dmftdmrg}.
In the Mott insulating phase DDMRG-DMFT  calculations  
are relatively easy to perform because there is no sharp structure
to resolve around $\omega=0$.
Thus we can obtain very accurate results for the Hubbard  bands
in the spectral density and determined the one-particle gap.
Our numerical results agree remarkably well with strong-coupling
perturbation theory~\cite{nextpaper}.
We are currently investigating the intermediate-coupling regime,
where the Mott metal-insulator transition occurs~\cite{florianbook}, using
DDMRG-DMFT.

The DDMRG method presented here  can be generalised and applied to
a great variety of quantum impurity problems.
First, although we have discussed only the symmetric flat-band
SIAM, our method can readily be applied to an asymmetric 
SIAM~\cite{lma2} or more
complicated hybridisation functions.
Second, the method can be used for other Hamiltonians than the SIAM
(for instance, the Kondo model) provided that the transformation to a 
one-dimensional lattice configuration with short-range
interactions only is possible.
Third, contrary to NRG our method can resolve sharp
spectral structures at any frequency and thus could be very useful
for studying the dynamics of systems with resonances at finite frequency,
such as the Kondo model in magnetic field~\cite{rosch}.
Fourth, DDMRG can be applied to models with other
degrees of freedom than fermions. For instance, there are very efficient
density-matrix renormalisation methods to treat bosons~\cite{chunli}.
One can easily combine them with DDMRG to investigate a quantum
degree of freedom coupled to a dissipative environment such as
a phonon bath~\cite{leggett}.
Finally, our method can be generalised to problems with more than one 
impurity or more than one host band.
If the impurity sites are arranged in a one-dimensional lattice and the 
number of host bands do not exceed two,
the computational effort is probably comparable to that required for
the investigation of the SIAM presented here.
Therefore, one should be able to investigate the dynamics of
a two-channel Kondo problem or of
quantum dots and wires coupled to (up to) two leads~\cite{qdots}
without difficulty.
If the impurity sites form other structures such as a two-dimensional 
cluster or the number of host bands is greater than two
(which occurs for multi-channel Kondo problems or
in the dynamical cluster approximation~\cite{dca}),
DDMRG calculations of dynamic properties are also possible but 
the computational
effort could be substantially greater.
In summary, DDMRG provides a powerful and versatile approach for 
investigating the dynamics of quantum impurity problems.

\ack

We are deeply grateful to F Gebhard for suggesting the 
extension of DDMRG to quantum impurity problems
and for many stimulating discussions.
We also wish to thank T Pruschke, D Logan, T Costi and R Bulla
for helpful discussions regarding the present work.

\end{document}